\documentclass[fleqn,twoside]{article}
\usepackage{espcrc2}

\usepackage{graphicx}
\usepackage[figuresright]{rotating}
\usepackage{latexsym}

\newcommand{\AmS}{{\protect\the\textfont2
  A\kern-.1667em\lower.5ex\hbox{M}\kern-.125emS}}

\title{The mass composition of cosmic rays above $10^{17}$ eV*}

\author{A A Watson\address{
School of Physics and Astronomy, University of Leeds, 
Leeds LS2 9JT, UK\\
$^{\ast}$Based on talk at XIII ISVHECRI: Pylos, Greece, September 2004}}

\begin{document}

\begin{abstract}
  Our knowledge of the mass composition of cosmic rays is deficient at 
all energies above $10^{17}$ eV.  Here, systematic differences between 
different measurements are discussed and, in particular, it is argued that 
there is no compelling evidence to support the common assumption that 
vast majority of the cosmic rays of the highest energies are protons.  
Our knowledge of the mass needs to be improved if we are to 
resolve uncertainties about the energy spectrum.  Improvement is also 
needed for proper interpretation of data on the arrival direction 
distribution of cosmic ray.  
\end{abstract}

\maketitle

\section{The Scientific Motivation for Studying the Highest Energy 
Cosmic Rays}
Since the recognition in 1966, by Greisen and by Zatsepin and Kuzmin, 
that protons with energies above $4 \times 10^{19}$ eV would interact 
with the cosmic microwave radiation, there has been great interest in 
measuring the spectrum, arrival direction distribution and mass 
composition of ultra high-energy cosmic rays (UHECR), defined as those 
cosmic rays having energies above $10^{19}$ eV.  Specifically they 
pointed out that if the sources of the highest energy protons are 
universally distributed, there should be a steepening of the energy 
spectrum in the range from 4 to $10 \times 10^{19}$ eV.  This feature 
has become known as the GZK `cut-off' but the sharpness of the 
steepening expected depends on unknown factors such as the evolution 
and production spectrum of the sources.  If the UHECR are mainly Fe 
nuclei then the spectrum is also expected to steepen, but it is harder 
to predict the character of this feature as the relevant diffuse 
infrared photon field is poorly known: steepening is expected to set in 
at higher energy.

\section{Importance of mass composition for accurate estimates of the 
energy spectrum}
The energy spectrum of cosmic rays has frequently been inferred from observations of signals in arrays of scintillators or water-Cherenkov 
detectors (see Nagano and Watson \cite{nag00} for a review).  In 
practice, the energy is derived from a measurement of the detector 
response at (typically) 600 m from the shower axis using the results of 
detailed Monte Carlo calculations.  For example, in the most recent 
reports of spectra from the AGASA \cite{tak03} and Haverah Park 
groups \cite{ave03}, the QGSJET model of high energy interactions has 
been adopted.  Additionally, in the AGASA work, a study of the impact 
of different models was made and it was found, for example, using 
QGSJET98 in the Corsika propagation code, that the energy estimates 
for protons or iron nuclei differ by about 12\% near $10^{20}$ eV 
(with protons giving the higher energy), while with SIBYLL 1.6 the 
difference is 19\%, in the same direction and the overall energy 
estimates are about 5\% higher.  These two sources of systematic error 
in energy estimates---from mass and model---are essential to 
keep in mind when comparing data sets.  As the relevant centre of 
mass energy at $10^{20}$ eV is well beyond that anticipated at the 
LHC, it is clear that the systematic effect in the model is extremely 
difficult to quantify.

The other method that has been used to measure primary energies 
is the fluorescence technique in which photons emitted by excited 
N$_{2}$ are detected.  With this method, it is possible to make a 
calorimetric estimate of the large fraction of the primary energy 
that is transferred into ionisation as the shower particles cross 
the atmosphere.  The estimate of this part of the primary energy loss 
is relatively model independent with, as first discussed by Linsley \cite{lin83}, and more recently by Song et al.\ \cite{son00},   
a correction of around 10\% being required for energy that does not 
go into atmospheric ionisation.  The most recent journal presentation 
of a fluorescence spectrum by the HiRes group is one in which monocular 
data from two detectors are reported \cite{abb04}.  Note that the 
correction for missing energy is mass dependent, with an energy estimate 
about 5\% higher being made if the primaries are assumed to be Fe nuclei.

When making comparisons of spectra, a log-log plot of J (the differential intensity) against energy, E, has significant advantages over the 
JE$^{3}$ vs. E plots commonly adopted in recent years.  In 
particular,  propagators of the latter style (including the present 
author) almost always ignore the fact that the error bars on such a 
plot should be shown as diagonal lines and that the uncertainty in 
energy (often justifiably omitted in the x-direction because of the bin 
size) should be added in quadrature to the uncertainty in intensity, as 
it comes in as the third power.   The JE$^{3}$ vs. E representations 
do the data a disservice as the incorrect error assignments serve to 
give an erroneous impression of the level of incompatibility between 
the data sets.  For example, the differences between the AGASA and HiRes 
spectra could largely be resolved, except perhaps at the very highest 
energies, if one or other of the energy estimates had a systematic 
error of only $\sim 20-30$\%.  Some movement towards reconciliation would arise if the primary particles were iron nuclei. 

\section{The mass of UHECR}
Knowledge about the mass of primary cosmic rays at energies above 
$10^{17}$ eV is rudimentary.  Different methods of measuring the 
mass give different answers and the conclusions are dependent, in all 
cases, upon the model calculations that are assumed.  Results from some 
of the techniques that have been used are now described and the 
conclusions drawn reviewed.  Some of these techniques will be applicable 
with the Pierre Auger Observatory \cite{aug04}.  

\subsection{The Elongation Rate}
\begin{figure}[htb]
\begin{center}
\vspace{-5mm}
\includegraphics[scale=1.07]{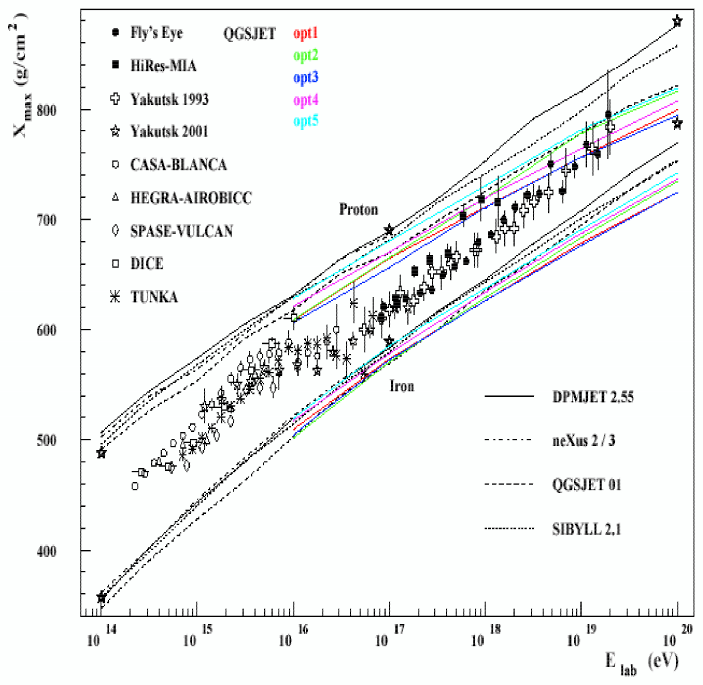}
\end{center}
\vspace{-8mm}
\caption{The depth of maximum, as predicted using various models, 
compared with measurements.  The predictions of the five modifications 
of QGSJET, discussed in [9], from which this diagram is taken, lie 
below the dashed line that indicates the predictions of QGSJET01.}
\label{fig1}
\end{figure}

The elongation rate is the term used to describe the rate of change of 
depth of shower maximum with primary energy.  The term was introduced by 
Linsley \cite{lin77} and, although his original approach has been, to 
some extent, superseded by the results of detailed Monte Carlo studies, 
the concept remains useful for organising and describing data 
and calculations.  In figure~\ref{fig1} there is a summary of 
measurements of the 
depth of maximum together with predictions from a variety of model 
calculations reproduced from \cite{zha03}.  It is clear that, if 
certain models are correct, one might infer that the primaries 
above $10^{19}$ eV are dominantly protons but that others are 
indicative of a mixed composition.  
In particular, the QGSJET set of models (basic QGSJET01 and the 5 
options discussed in \cite{zha03}) and the Sibyll 2.1 model force 
contrary conclusions.  Note that the experimental data on X$_{max}$ 
do not yet extend beyond 25 EeV.

\subsection{Fluctuations in Depth of Maximum}
Further insight is expected to come from the magnitude of the fluctuation 
of the position of depth of maximum, X$_{max}$.  If a group of showers 
is selected within a narrow range of energies, then fluctuations about 
the mean of X$_{max}$ (or in a parameter that is closely related to 
X$_{max}$ such as the steepness of the lateral distribution function 
or the spread of muon densities) are expected to be larger for protons 
than for iron nuclei.  A recent study applying this idea to X$_{max}$ 
has been reported by the HiRes group \cite{abbas} for 553 events 
above 10 EeV. 

\begin{figure*}[htb]
\begin{center}
\includegraphics[scale=0.95]{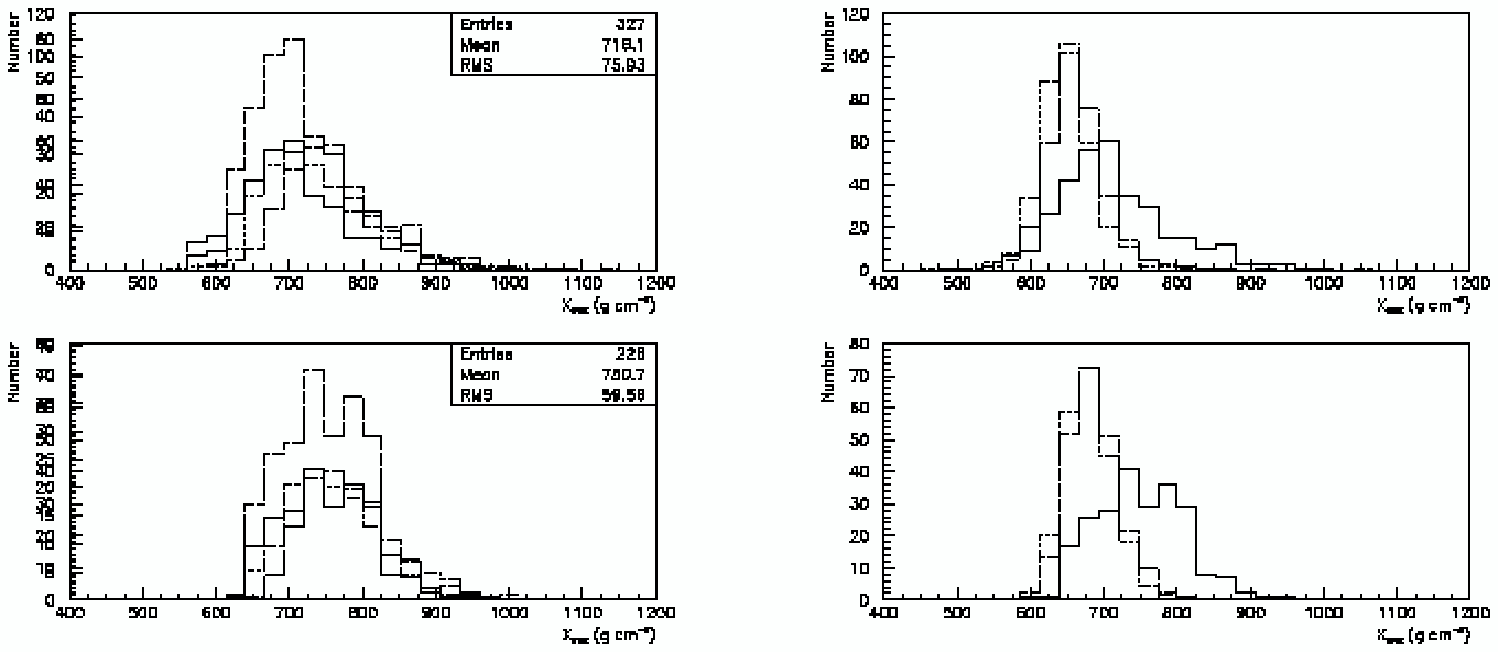}
\end{center}
\vspace{-8mm}
\caption{Presentation in [10] of the distribution of the measured 
values of X$_{max}$ for the energy intervals 1 to 2.5 EeV and 2.5 to 25 EeV.  
The higher energy range is in the lower plots and the data (solid lines) 
are compared with the QGSJET model (dashed) and the Sibyll model (dotted).}
\label{fig2}
\end{figure*}

In \cite{abbas} it is argued that the fluctuations are so large that a 
large fraction of light nuclei must be present in the primary 
beam (figure~\ref{fig2}): this conclusion is independent of whether Sibyll or 
QGSJET models are used.  Clearly an understanding of the tails on the 
high-X$_{max}$ side of the histograms is crucial to the conclusion that 
they are best described by `a predominantly light composition'.  Of the 
553 events in the data set, detailed atmospheric information is available 
for 419: the rest (134) occurred during `good weather' and were 
analysed assuming average atmospheric conditions.  This assumption may 
not represent reality, as it is possible that the atmosphere deviates from 
the standard conditions from night to night and even during a night 
of observation.  This view is strengthened by the results of balloon 
flights made from Malarg\"{u}e \cite{kei00}, which have shown that the atmosphere changes in a significant way both diurnally and seasonally.  If 
a standard atmosphere is used, some of the fluctuations observed in 
X$_{max}$ may be attributed incorrectly to shower, rather than to 
atmospheric, variations.  It would be instructive to see the distributions 
in X$_{max}$ for the two weather groups.

Questions about the interpretation are also raised when the Monte 
Carlo analysis of uncertainties in X$_{max}$ and the fluctuations in 
X$_{max}$ are considered (Perrone \cite{per04}).  He finds that at 
distances beyond 20 km, there are significant systematic shifts in the 
X$_{max}$ values derived and in the spread of the X$_{max}$ values.  At 
20 km, X$_{max}$ is estimated to be 60 g cm$^{-2}$ deeper in the 
atmosphere, on average, than reality and the fluctuations in X$_{max}$ for 
iron nuclei are considerable with $\sigma \sim 100$ g cm$^{-2}$.  These 
factors act in such a way as to suggest that the elongation rate 
reported by HiRes (and presumably also by Fly's Eye) may have been systematically over-estimated and that the fluctuations in X$_{max}$ are 
not due entirely to protons. Thus, it may be premature to draw 
conclusions about the presence of light nuclei from the analyses of 
fluctuations as presented so far.

A further issue of some concern is the quality of events selected for the 
HiRes analysis.  In \cite{abbas} it is stated that events were selected 
when $\chi^{2}$ per degree of freedom for the fit of the longitudinal development was less than 20.  This is a rather loose cut.  Furthermore, 
the uncertainties in X$_{max}$ are required to be less than 200 g cm$^{-2}$ 
for the fit made using both Eyes (as compared with 400 g cm$^{-2}$ for 
each Eye).  It is hard to reconcile these numbers with the resolution 
of 30 g cm$^{-2}$ claimed for the measurements from Monte Carlo studies 
and it seems reasonable to question whether all of the events in the 
tails in the data of 
figure~\ref{fig2} arise from the presence of light primary particles. 

\subsection{Mass from muon density measurements}
It is well known that a shower produced by an iron nucleus will contain 
a greater fraction of muons at the observation level than a shower of the 
same energy created by a proton primary.  Many efforts to derive the mass
spectrum of cosmic rays have been attempted using this fact over the full 
range of air shower observations.  However, although the differences are predicted to be relatively large (on average there are $\sim70$\% more 
muons in an iron event than a proton event), there are large fluctuations 
and, again, there are differences between what is predicted by 
particular models.  Thus, the QGSJET model set predicts more muons 
than the Sibyll family, the difference arising from different predictions 
as to the pion multiplicities produced in nucleon-nucleus and 
pion-nucleus collisions that in turn arise from differences in the 
assumptions about the parton distribution within the nucleon \cite{alv02}.  
In a contribution to these proceedings, Shinosaki has described the data 
on muons signals from the AGASA array.  There are 129 events above 
$10^{19}$ eV, of which 19 have energies greater than $3 \times 10^{19}$ eV.  
Measurements of muon densities at distances between 800 and 1600 m were 
used to derive the muon density at 1000 m with an average accuracy of 40\%.  
This quantity is compared with the predictions of model calculations.  The difference between the proton and iron predictions is small, especially when fluctuations are considered.  The AGASA group conclude that above $10^{19}$ eV 
the fraction of Fe nuclei is $< 40$\% at the 90\% confidence level.  
In my view, the 5 events above $10^{20}$ eV for which such measurements 
are possible, are fitted as well by iron nuclei as by protons.  Further, the conclusions are sensitive to the model used: as the Sibyll model predicts 
fewer muons than the QGSJET model, higher iron fractions would have been 
inferred had that model been adopted.  At this meeting, Ostapchenko has discussed changes to the QGSJET model that will reduce the number of muons expected for a proton primary (see also Engel \cite{eng04}).  The magnitude 
of the effect is not yet clear but it is in such a direction as to raise 
the fraction of iron nuclei at the energies in question.  

At lower energies, there are muon data from the Akeno array and from 
AGASA \cite{hay95}.  Different analyses have been made of these.  The 
AGASA group claim that the measurements are consistent with a mass 
composition that is unchanging between $10^{18}$ and $10^{19}$ eV.  
Another interpretation is discussed below.  

\subsection{Mass estimates from the lateral distribution function}
The rate of fall of particle density with distance from the shower axis 
provides a further parameter that can be used to deduce the mass 
composition.  Showers with steeper lateral distribution functions (LDFs) 
than average will arise from showers that develop later in the atmosphere, 
and vice versa.  A detailed measurement of the LDFs of showers produced by primaries of energy greater than $10^{17}$ eV was made at Haverah Park 
using a specially constructed `infilled array' in which 30 additional 
water tanks of 1 m$^{2}$ were added on a grid with spacing of 150 m.  When 
the experimental work was completed in 1978, the data could not be fitted 
with the 
shower models then available for any reasonable assumption about the 
primary mass.  Recently \cite{ave02}, these data have been re-examined 
using the QGSJET98 model.  The choice of model was justified by showing 
that it adequately described data on the time spread of the Haverah 
Park detector signal over a range of zenith angles and distances near the 
core ($<500$ m).  Here the difference predicted between the average proton 
and iron shower is only a few nanoseconds and the fit achieved is good.  
Density data were fitted by a function 
$\rho$(r) $\sim$ r $^{(\eta + r/4000)}$, where $\eta$ is the steepness 
parameter.  The spread of $\eta$ was compared with predictions for 
different primary masses.  The proton fraction, assuming a proton-iron 
mixture, is found to be independent of energy in the range 
$3 \times 10^{17}$ to $10^{18}$ eV and is ($34\pm2$) \%.  When this 
fraction is evaluated with QGSJET01, in which a different treatment of diffractive processes is adopted from that in QGSJET98, then the fraction 
rises to 48\%.  The fraction is larger because the later model predicts 
shower maxima that are higher in the atmosphere and accordingly, to 
match the observed fluctuations, the proton fraction must be increased.  
The difference in the deduced ratio thus has a systematic uncertainty from 
the models that is larger than the statistical uncertainty.  Although the necessary analysis has not been made, it is clear that the Sibyll 2.1 model would be consistent with an even smaller fraction of protons.  

A similar analysis has been carried out using data from the Volcano Ranch 
array.  As with the Haverah Park information, no satisfactory 
interpretative analysis was possible when the measurements were completed.  
Using 366 events for which Linsley left detailed information, and QGSJET98, 
the fraction of protons is estimated as 
(11 $\pm$ 5(stat) $\pm$ 12 (sys))\% at $\sim$ 1 EeV \cite{dov03}.  With 
QGSJET01 the flux of protons would increase to $\sim$ 25\%, indicating 
again that model uncertainties remain a serious barrier to interpretation.

\subsection{Mass from the thickness of the shower disk}
The particles in the shower disc do not arrive at a detector 
simultaneously, even on the shower axis.  The arrival times are spread 
out because of geometrical effects, velocity differences, and because of 
delays caused by multiple scattering and geomagnetic deflections.  The 
first particles to arrive (except very close to the shower axis) are the 
muons as they are scattered rather little and geometrical effects dominate.  
At Haverah Park four detectors, each of 34 m$^{2}$, provided a useful tool 
for studying how the thickness of the shower disc depends upon the development 
of the cascade.  Recently, an analysis of 100 events of mean 
energy $\sim10^{19}$ eV has shown that the magnitude of the risetime 
is indicative of a large fraction ($\sim$80\%) iron nuclei at this 
energy \cite{avm03}.  This type of study will be considerably extended with the Pierre Auger Observatory in which the photomultipliers within each water tank are equipped with 25 ns flash ADCs.

\subsection{Summary of data on primary mass above 0.3 EeV}
\begin{figure}[htb]
\begin{center}
\includegraphics[scale=0.85]{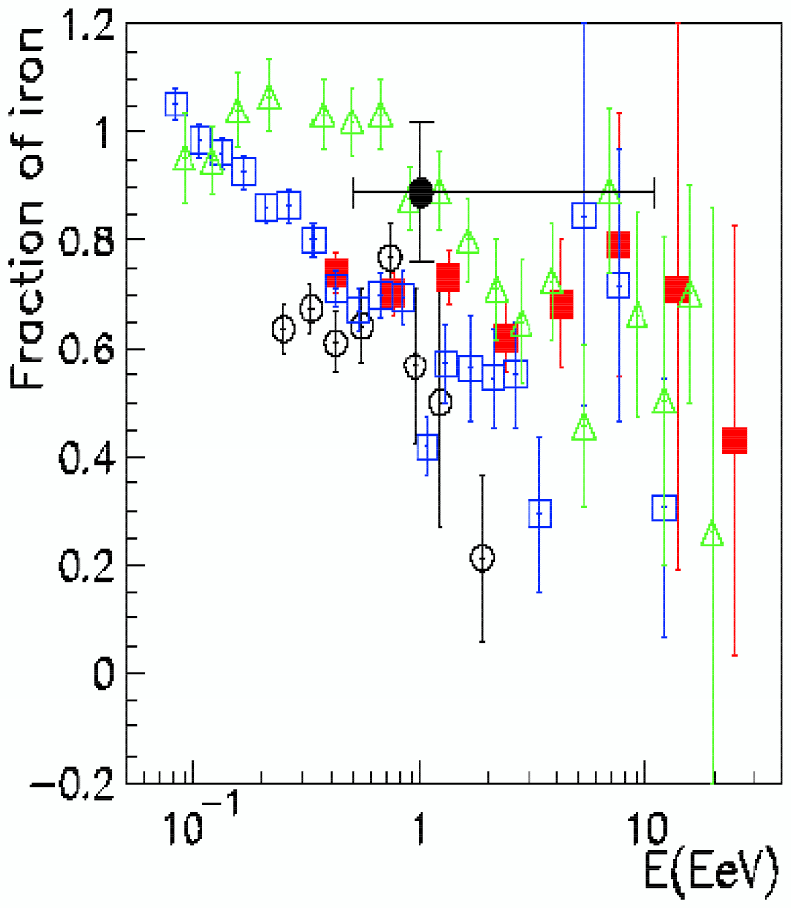}
\end{center}
\vspace{-8mm}
\caption{The fraction of Fe nuclei as a function of energy as reported from various experiments.  Fly's Eye ($\bigtriangleup$), AGASA A100 (\protect\rule{2mm}{2mm}), AGASA A1 ($\Box$), Haverah Park ($\bigcirc$) and Volcano Ranch ($\bullet$) (Figure taken from [17]).}
\label{fig3}
\end{figure}

In figure~\ref{fig3}, taken from \cite{dov03}, the results taken from various 
reports of the Fe fraction are shown.  It is disappointing that the data from 
Volcano Ranch and from Haverah Park are not in better agreement as a 
similar quantity, the lateral distribution function of the showers, was 
measured at each array and the same model - QGSJET98 - was used for interpretation, although with different propagation codes (AIRES and CORSIKA respectively).  We cannot explain this difference: at $10^{18}$ eV the 
estimates of the fraction of Fe are separated by over 2 
standard deviations.

In figure~\ref{fig3}, data from the Akeno/AGASA and the Fly's Eye experiments are also shown.  The Akeno/AGASA groups measured the muon densities in 
showers, normalised at 600 m.  The energy thresholds for Akeno and AGASA 
were 1 and 0.5 GeV respectively.  The Fly's Eye data are deduced from measurements of the depth of shower maximum.  In an effort to reconcile differing claims made by the two groups of the trend of mass composition 
with energy, Dawson et al.\ \cite{daw98} reassessed the 
situation using a single model, SIBYLL 1.5 on both data sets.  SIBYLL 1.5 
was an early version of the
SIBYLL family that evolved to SIBYLL 1.6 and 1.7.  It is the estimates 
of the Fe fractions from \cite{daw98} that are shown in figure~\ref{fig3}.  
There are major discrepancies between these estimates and between those from 
Volcano Ranch and Haverah Park.  However, the predictions of the muon 
density and of the depth of shower maximum made with the version of 
SIBYLL used in \cite{daw98} differ significantly from those that would be derived now using QGSJET98 or 01 (or with the later SIBYLL version, 2.1).  
We now discuss this point in some detail.

An extremely useful set of comparisons of the predictions from 
SIBYLL 1.7 and 2.1 with those from QGSJET98 has been given in \cite{alv02}.  
We understand that SIBYLL 1.6 and SIBYLL 1.7 differ only in that the neutral pions were allowed to interact in the latter model and it is not believed 
that this will make a serious difference to the predictions at energies 
below $10^{19}$ eV \cite{sta04}.  Therefore, in what follows, we regard 
the SIBYLL 1.7 and the QGSJET98 differences as being identical to those 
that exist between SIBYLL 1.6 (or 1.5) and QGSJET98, for which no similar comparisons are available.  It is convenient to compare conclusions 
at $10^{18}$ eV.  More detailed cross-checks, over a range of energies, would require more extensive knowledge of features of the Fly's Eye and Akeno/AGASA systems than we possess.

Turning first to the data from the depth of maximum, we note that at 
$10^{18}$ eV the measured value of X$_{max}$ is $\sim$675 g cm$^{-2}$, with 
an error that is less than the size of the data point ($<$ 10 g cm$^{-2}$).  
The predictions for proton primaries made with SIBYLL 1.7 and 
QGSJET98 are 760 and 730 g cm$^{-2}$ respectively \cite{alv02}.  Thus, 
a mass composition less dominated by Fe is favoured compared with 
the $\sim90$\% estimated in \cite{daw98}.  The choice of SIBYLL 2.1 would 
alter this argument rather little as the predicted depth at 
$10^{18}$ eV is 740 g cm$^{-2}$ \cite{alv02}.  Further study of this 
matter could be made but the data from Fly's Eye has now been 
complemented by data from the HiRes stereo system \cite{abbas} (although 
the differences between the data sets have not been explained) and there will also be data from the Auger instruments.

A qualitative statement about the shift expected in the Fe fraction, as estimated from the measurement of muon densities at 600 m, from changes in the model can be made using information in \cite{alv02}.  Although the 
calculations do not exactly match the energies of the Akeno/AGASA 
measurements ($>$ 0.3 GeV is computed and $>$ 0.5 GeV measured), ratios 
between the predictions of different models are not strongly dependent 
upon energy threshold.  What is of importance is the ratio of the number 
of muons predicted, at $10^{18}$ eV, for SIBYLL 1.7, SIBYLL 2.1 and 
QGSJET98.  At $10^{18}$ eV, the numbers are in the ratios 1: 1.17: 1.44.  
The difference in muon number between SIBYLL 1.7 and QGSJET98 is 
comparable to that expected between proton and Fe primaries ($\sim$50\%, 
but also model dependent).  It is clear that the more recent models, if 
applied to the Akeno/AGASA data after the manner of the analysis of 
\cite{daw98}, would lead to a significant reduction in the predicted fraction of Fe nuclei.  To pursue this further would require knowledge of the predicted densities at 600 m, information that is presently lacking.  We note that the shift in the Fe fraction from the muon data is probably substantially larger than it is when using the data on X$_{max}$.  

We are not able to use the information reported from the HiRes-MIA 
experiment \cite{abu00} in which muons and X$_{max}$ were observed 
simultaneously.  As with Akeno/AGASA, the muon density at 600 m was 
determined.  The problem we have is that while the papers describe the 
data as being consistent with a mass composition that becomes lighter 
with energy, this appears, on close scrutiny of figures~\ref{fig1} 
and ~\ref{fig2} of 
\cite{abu00}, to be true only for the X$_{max}$ data.  The muon data, 
which are compared with predictions of QGSJET98, look to be consistent 
with a constant and heavy mass from $5 \times 10^{16}$ to beyond 
$10^{18}$ eV.  It would be very interesting to establish that the 
same model gives different predictions for the mass variation with 
energy for different measured quantities: this might lead to further
understanding of the models, or of the systematic errors in measurements 
of X$_{max}$, as discussed above.

This discussion is intended to demonstrate the difficulties with which 
one is faced with when trying to compare data.  Measurements from 
different experiments are rarely analysed contemporaneously and the 
shifts in the inferences from the use of different models can be substantial.
  
\section{Possibilities of Identifying Photon and Neutrino Primaries}

\subsection{Super-heavy relic particles}

An idea to explain the UHECRs that have been reported beyond 100 EeV is 
that super-heavy relic particles with masses of $\sim10^{12}$ GeV, produced 
in the early Universe, may decay to produce UHECR \cite{ber97}.  
While details of the fragmentation of these particles remains a matter 
of debate, it is generally accepted that the resulting UHECR beam would 
contain copious fluxes of neutrinos and photons.  
 
\subsection{Limits to the fraction of photon primaries}

It is unlikely that the majority of the events claimed to be near 
$10^{20}$ eV have photons as parents as some of the showers seem to 
have normal numbers of muons, the tracers of primaries that are 
nuclei\footnote{This assumes that the photo-pion production 
cross-section behaves `normally'}, 
(see paper in these proceedings by Shinosaki).  It has been 
argued that the cascade profile of the most energetic Fly's Eye 
event \cite{bir95} is inconsistent with that of a photon primary \cite{hal95}.  However, in a recent paper by Risse et al.\ \cite{ris04} (and Risse, these proceedings) have used calculations made with the QGSJET01 and 
SIBYLL2.1 models to show that the profile of this event can be explained 
under the assumption of any baryonic primary between a proton and iron 
nucleus, and that the primary photon hypothesis, although not favoured, cannot be rejected. 

An alternative method of searching for photons has been developed 
using showers incident at very large zenith angles.  Deep-water tanks 
have a good response to such events out to beyond $80^{\circ}$.  At such 
angles the bulk of the showers detected are created by baryonic 
primaries but they are distinctive in that the electromagnetic cascade 
stemming from neutral pions has been almost completely suppressed by the 
extra thickness of atmosphere penetrated.  At $80^{\circ}$ the 
atmospheric thickness is $\sim$ 5.7 atmospheres at sea-level.  At Haverah 
Park, showers at such large zenith angles were observed and the shower disc 
was found to have a very small time spread.  A complication for the study 
of inclined showers is that the muons, in their long traversal of the atmosphere, are very significantly bent by the geomagnetic field.  A 
study of this has been made and it has been shown that the rate of 
triggering of the Haverah Park array at large angles can be predicted 
\cite{ave00}.  In addition, it was found that the energy of the 
primaries could be estimated with reasonable precision so that an 
energy spectrum could be derived.  The concept of using the known, and 
mass independent, spectrum deduced from fluorescence detectors to 
predict the triggering rate as a function of the mass of the primary has 
led to a demonstration that the photon flux at $10^{19}$ eV is less than 40\% 
of the baryonic component \cite{avm00}, a conclusion similar to that 
of the AGASA group, made by searching for showers which have significantly 
fewer muons than normal \cite{shi02}.  

The flux of photons expected at $10^{19}$ eV from super-heavy relic 
particles has recently been reassessed \cite{ber04} and is expected to be 
lower than originally predicted.  This arises because of a more 
detailed examination of the fragmentation process.  A large flux of 
photons is now predicted at 100 EeV and can be sought in the Auger data 
using the method developed at Haverah Park.

\subsection{Neutrino Primaries:} Neutrino primaries may be detectable 
by studying very inclined showers.  This idea was first proposed by 
Berezinsky and Smirnov \cite{ber75} and was re-examined in the context 
of the Auger Observatory by Capelle et al.\ \cite{cap98}.  A discussion 
of the potential of the Pierre Auger Observatory to detect such events 
is given in the paper by K-H Kampert in the proceedings of this meeting.  
A neutrino can interact anywhere in the atmosphere with equal probability.  

\section{The Zatsepin effect}
One method capable of giving an estimate of the primary mass, that 
makes no assumptions about the particle physics at extreme energies, is 
that proposed by Zatsepin \cite{zat51} in 1951.  He pointed out that 
heavy nuclei would undergo photodisintegration in the radiation field of 
the sun.  This is the same process, but with the 2.7 K and IR radiation 
fields, that would affect the spectrum of Fe nuclei at the very highest energies.  The process in the solar photon field is important for $^{56}$Fe 
nuclei at energies of around $10^{18}$ eV and the resulting 
fragments ($^{55}$Fe and a neutron) would produce showers at roughly the same 
time and at nearly the same place in the atmosphere.  In a later 
calculation, Gerasimova and Zatsepin \cite{zat51} overlooked the effect 
of the interplanetary magnetic field and predicted a separation of 
fragments of only a few 100 metres.  When the effect of the interplanetary 
field was included the core separation was estimated to be many 
kilometres\footnote{During the Pylos meeting, G T Zatsepin credited 
M Shapiro with pointing out the effect of the interplanetary magnetic 
field in discussions during the 5th International Cosmic Ray 
Conference (Guanajauto 1955)}.  
The importance of Zatsepin's idea is that the ratio of 
the sizes of the two correlated showers would be proportional to the mass 
of the fragmented primary.  No assumptions about particle physics 
are needed.

In view of the scale of the Pierre Auger Observatory ($\sim$ 31 km radius) 
and the improved knowledge of the interplanetary field, a further study 
was made of this effect \cite{ger60}.  However, it was found that core 
separations of less than 10 km (a scale relevant to the Auger surface array and 
to AGASA) were infrequent (0.3 per year on Auger) and hard to detect, as the 
optimum energy of about $6 \times 10^{17}$ eV is rather low.  However, 
there is a stronger signal for separations of $\sim10^{3}$ km, the 
approximate distance between two sites being considered for the northern 
Auger Observatory in Utah and South Eastern Colorado: this prospect may be worthy of more detailed scrutiny.  A model-independent measurement of the mass of primary cosmic rays is highly desirable.  This beautiful idea may 
provide a solution.
 
\section*{Conclusions}
To make full use of forthcoming information on the energy spectrum and 
arrival direction distribution at the highest energies, and to interpret 
what already exists, it is necessary to improve our knowledge of the 
mass of the cosmic rays above $10^{19}$ eV.  Such evidence as there is 
does not support the widely adopted assumption that all of these cosmic 
rays are protons: there may be a substantial fraction of iron nuclei 
present, even at $10^{20}$ eV.  Photons do not appear to dominate 
at $10^{19}$ eV.

\section*{Acknowledgements}

I would like to thank the organisers of the Conference for inviting me 
to give the talk on which this paper is based and for financial support.  
Lively discussions with Tere Dova sharpened some of the arguments 
presented in \S3.6 and I gratefully acknowledge illuminating discussions 
with Sergei Ostapchenko and Ralph Engel about the new QGSJET model.  Work 
on UHECR at the University of Leeds is supported by PPARC, UK


\begin{thebibliography}{99}

\bibitem{nag00}Nagano, M, and A A Watson, Rev Mod Phys 72 689 2000
\bibitem{tak03}Takeda, M. et al.,  Astropart Phys 19 447 2003 and astro-ph/0209442
\bibitem{ave03}Ave, M. et al.,  Astropart Phys 19 47 2003
\bibitem{lin83}Linsley, J., Proc 18th ICRC, Bangalore, 12 144 1983
\bibitem{son00}Song, C., et al.,  Astropart Phys 14 7 2000
\bibitem{abb04}Abbasi, R.U., et al., Phys Rev Lett 92 151101 2004
\bibitem{aug04}Auger Collaboration: Abraham, J., et al., NIMA 523 50 2004
\bibitem{lin77}Linsley, J.,  Proc 15th Int Cos Ray Conf (Plovdiv) 12 89 1977
\bibitem{zha03}Zha, M., J Knapp and S Ostapchenko, Proc 28th Int Cos Ray Conf (Tsukuba) 2 515 2003
\bibitem{abbas}Abbasi, R.U., et al., HiRes Collaboration : astro-ph 0407622
\bibitem{kei00}Keilhauer, B., et al., Proc 28th Int Cos Ray Conf (Tsukuba) 2 879 2000 and Astroparticle Physics (in press)
\bibitem{per04}Perrone, L., for the Auger Collaboration, Proceedings of the CRIS meeting, Catania, June2004 (in press)
\bibitem{alv02}Alvarez-Mu\~{n}iz, J., et al., Phys. Rev. D66 033011 2002 and astro-ph/0205302 2002
\bibitem{eng04}Engel, R, 3$^{rd}$ International Workshop on UHECR, Leeds, 
July 2004:\\                                                                              www1.ast.leeds.ac.uk/\\ $\sim$workshop/workshoprogramme.html
\bibitem{hay95}Hayashida, N., et al., J Phys G 21 1101 1995
\bibitem{ave02}Ave, M., et al., Astroparticle Physics 19 61 2003 and      astro-ph/0203150
\bibitem{dov03}Dova,  M. T., et al., Proc 28th Int Cos Ray Conf (Tsukuba) 1 377 2003 and Astropart Phys 21 597 2004
\bibitem{avm03}Ave, M., et al., Proc 28th Int Cos Ray Conf (Tsukuba) 1 349 2003 
\bibitem{daw98}Dawson, B.R., R Meyhandan and K M Simpson, Astropart Phys 9 331 1998
\bibitem{sta04}Stanev, T., private communication, February 2004
\bibitem{abu00}Abu-Zayyad, T., et al., Phys Rev Lett 84, 4276 2000\\
        Abu-Zayyad, T., et al., Ap J 557 686 2001
\bibitem{ber97}Berezinsky, V., Kachelreiss, M. and Vilenkin, A., Phys Rev Letters 22 4302 1997\\
         Benakli, K., Ellis, J. and Nanopolous, D.V., Phys Rev D59 047301 1999\\
         Birkel, M and Sarkar, S., Astroparticle Physics, 9, 297 1998\\
         Chisholm, J.R. and E W Kolb, astro-ph/0306288, submitted to Phys Rev D\\
         Chung, D J H., E W Kolb and A Riotto, Phys Rev Letters 81 4048 1998\\
         Rubin, N. A., M Phil Thesis, University of Cambridge, 1999\\
         Sarkar, S., and Toldra, R., Nuclear Physics B 621 495 2002
\bibitem{bir95}Bird, D.J., et al., ApJ 441 144 1995
\bibitem{hal95}Halzen, F., et al. Astropart Phys 3 151 1995
\bibitem{ris04}Risse, M., et al., Astroparticle Physics 21 479 2004
\bibitem{ave00}Ave, M., et al., Astropart Phys 14 109 2000 and astro-ph/0003011
\bibitem{avm00}Ave, M., et al., Phys Rev Letters 85 2244 2000 and astro-ph/0007386
\bibitem{shi02}Shinosaki, K., et al., Astrophysical Journal 571 L117 2002
\bibitem{ber04}Berezinsky, V.S., Proceedings of the CRIS meeting, Catania, June 2004 (in press)
\bibitem{ber75}Berezinsky, V. S., and A Yu Smirnov, Astrophys Space Sci 32 461 1975
\bibitem{cap98}Capelle, K. J., et al. Astropart Phys 8 321 1998
\bibitem{zat51}Zatsepin, G., Dokl. Akad. Nauk. SSSR, 80, 577 (1951).
\bibitem{ger60}Gerasimova, N.M., and Zatsepin, G.T., Soviet Phys. (JETP), 11, 899 (1960).
\bibitem{med99}Medina-Tanco, G.A., and Watson, A. A., Astroparticle Physics, 10, 157 (1999)

\end{thebibliography}
\end{document}